# Prediction of experimental excited-state absorption spectra by using vibronic transition calculations: Its practical application to a π-conjugated molecule

Joonyoung F. Joung[1*] and Sungnam Park[2*]

[1]Department of Chemistry, Kookmin University, Seoul, 02707, Korea

[2]Department of Chemistry, Korea University, Seoul, 02841, Korea

AUTHOR INFORMATION

**Corresponding Authors**

* E-mail: jjoung@kookmin.ac.kr (J. F. Joung), spark8@korea.ac.kr (S. Park).

## Abstract

Accurate theoretical prediction of excited-state absorption (ESA) spectra remains challenging. Although considerable progress has been made in computational methods for ESA spectroscopy, their practical application to the interpretation of transient absorption and photoinduced absorption spectra is still limited. In this study, we present a practical approach for calculating molecular ESA spectra by combining vibronic transition calculations with routine density functional theory (DFT) and time-dependent DFT (TDDFT) calculations available in standard quantum chemistry packages. Using azulene as a model system, we demonstrate that the proposed method successfully reproduces the vibronic features and band positions observed in the experimental ESA spectrum. This approach therefore provides a practical framework for assigning ESA bands arising from transitions between low-lying excited states of molecules. The method delivers the vibronic band shapes and peak positions that are needed to assign ESA bands directly from these standard calculations, without requiring the excited-state-to-excited-state transition moments that determine absolute intensities.

**TOC graphics**

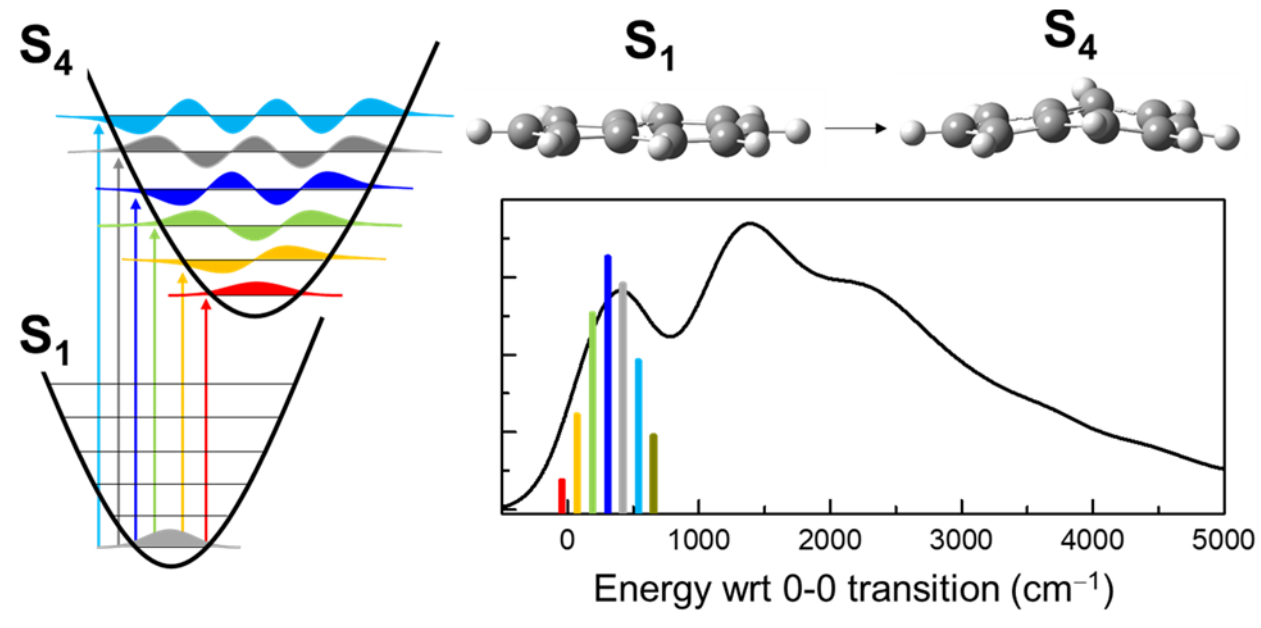

## 1. Introduction

Light-driven molecular processes such as photocatalysis,[1-2] light harvesting,[3] and excited-state proton transfer reactions,[4-7] together with the spectroelectrochemical properties of molecular systems, have been effectively studied by conducting transient absorption (TA) and steady-state photo-induced absorption (PIA) experiments.[8-10] TA and PIA spectra comprise the ground state bleaching (GSB), stimulated emission (SE), and excited-state absorption (ESA) contributions. The signs of the GSB and SE contributions are opposite to that of the ESA contribution, which sometimes restricts the analysis of TA and PIA spectra. Additionally, in light-driven molecular processes, such as electronic relaxation dynamics, resonance energy transfer, and photoinduced electron transfer, the ESA contribution may result from various highly excited and charged states. Density functional theory (DFT) and time-dependent DFT (TDDFT) methods can be utilized to accurately predict a molecule's optimized geometries, vibrational modes, electronic energies of the ground and excited states, and absorption and emission spectra. To analyze the TA and PIA, the GSB and SE spectra, which are considered equivalent to absorption and emission spectra, are easily obtained performing quantum chemical calculations.[11-16] However, reliable calculations of ESA spectra have been quite challenging.[17-18]

In the TDDFT, the linear response functions provide the transition energies and oscillator strengths between an electronic ground state ($S_0$) and higher-energy excited states ($S_n$, $n > 0$).[19-25] However, the corresponding transition properties between two electronic excited states, which allow the calculation of ESA spectra, cannot be obtained by the linear response functions.[26] Nevertheless, theoretical developments have rendered possible the calculations of ESA spectra; among these developments are the second linear response (SLR),[27-28] quadratic response (QR),[29-30] and real-time (RT) TDDFT methods.[31-33] The SLR-TDDFT method is essentially a sequential TDDFT calculation.[27-28] In particular, the optimized structure and molecular orbitals of a molecule in an excited state are first calculated using the TDDFT method; subsequently, the transition energies and oscillator strengths for the transitions from an excited state to higher-energy excited states are obtained by performing a second TDDFT calculation. In the QR-TDDFT method, quadratic response functions enable us to directly calculate two photon processes such as two-photon absorption and ESA spectra.[29-30] One advantage of the QR-TDDFT method over other methods is that only the molecular structure in its ground state needs to be optimized. However, the time-dependent exchange-correlation potential reduces the accuracy of this approach by red-shifting the transition energies calculated from the initial excited state of interest.[28] The RT-TDDFT method relies on simulations of the absorption spectrum of a molecule obtained by a Fourier transformation of the electron density fluctuation in the time domain to the frequency domain.[34] Recently, the RT-TDDFT method has been employed to calculate the ESA spectra of molecules in their excited states.[31-33] In fact, both SE and ESA spectra can be successfully estimated by the RT-TDDFT method, because the fluctuation of a molecule's dipole moment in its excited states is calculated under an external field. One disadvantage of the RT-TDDFT method is that the final electronic states associated with the ESA transitions cannot be accurately assigned. Furthermore, another challenge one faces when employing the RT-TDDFT method has to do with the unpredictable shifting of the transition energies. Notably, the methods described above are under development and have limitations for practical use in interpreting TA and PIA spectra. More recently, transient absorption spectra have also been simulated from first principles by combining nonadiabatic excited-state dynamics with an explicit probe step.[35-37]

In this work, we demonstrate a practical method for obtaining the ESA spectra of target molecules using vibronic transition calculations, without resorting to specialized response-theory implementations. It has been well established that the absorption and emission spectra of a molecule are readily obtained by calculating the vibronic transition between two electronic states in the framework of the Franck–Condon principle when the potential energy surfaces (PESs) of the corresponding electronic states are known.[11-16] Likewise, the vibronic band shapes and peak positions of the ESA spectra between any two electronic states can be obtained from vibronic transition calculations as long as the PESs of the electronic excited states are known, because the Franck–Condon factors depend only on the PESs and not on the electronic transition moment. The intensities, in contrast, require the transition moment between the two excited states, which lies beyond the scope of the present practical approach. In principle, any TDDFT methods can be used to obtain the PESs of electronic excited states. This idea has been demonstrated by Gierschner and coworkers.[29] They first used the QR TDDFT method to calculate the PESs of electronic excited states and then obtained the ESA spectra by using vibronic transition calculations. To demonstrate our method, we used azulene as a model molecule. We calculated the absorption, emission, and ESA spectra of azulene by using vibronic transition calculations in conjunction with the TDDFT method and directly compared them with the experimentally measured spectrum.

## 2. Theoretical background

The theoretical justification of the performance of vibronic transition calculations has been described elsewhere in detail.[13-14, 16] In this section, the general mathematical framework of vibronic transition calculations is briefly described. The absorption transitions, which are defined as the rate of energy absorption by a single molecule per unit radiant energy density, are given by[38]

$$\sigma_{\mathrm{abs}}(\omega) = \frac{4\pi^2\omega}{3c} \sum_{\mathrm{f}} \left| \langle \Psi_{\mathrm{i}} \mid \mu \mid \Psi_{\mathrm{f}} \rangle \right|^2 \delta(E_{\mathrm{f}} - E_{\mathrm{i}} - \hbar\omega) \tag{1}$$

where $\omega$ is the resonance frequency, $\langle \Psi_{\mathrm{i}} | \mu | \Psi_{\mathrm{f}} \rangle$ is the transition dipole moment integral, and $\Psi$ represents the wavefunction of the specified electronic state. The absorption transitions defined by Eq. (1) are usually convoluted with a Lorentzian or a Gaussian envelope to simulate a homogeneous or inhomogeneous broadening, respectively. In the Born–Oppenheimer approximation, the wavefunction and the electric dipole moment can be separated into their nuclear and electronic components. The transition dipole moment can be calculated by:

$$\begin{aligned} \langle \Psi_{\mathrm{i}} | \mu | \Psi_{\mathrm{f}} \rangle &= \langle \psi_{\upsilon} \psi_{\mathrm{S}_n} | \mu_{\mathrm{e}} | \psi'_{\mathrm{S}_{n'}} \psi'_{\upsilon'} \rangle + \langle \psi_{\upsilon} \psi_{\mathrm{S}_n} | \mu_{\upsilon} | \psi'_{\mathrm{S}_{n'}} \psi'_{\upsilon'} \rangle \\ &= \langle \psi_{\upsilon} | \mu_{\mathrm{if}} | \psi'_{\upsilon'} \rangle = \mu_{\mathrm{if}} \langle \psi_{\upsilon} | \psi'_{\upsilon'} \rangle \end{aligned} \quad (2)$$

The second term in Eq. (2) vanishes owing to the orthogonality condition of electronic wavefunctions. Because no analytical expression exists for Eq. (2), some approximations are necessary to calculate the value of the integral. The transition dipole moment can be expanded in a Taylor series of the vibrational normal coordinates of the final electronic state ($Q'$):

$$\mu_{\mathrm{if}} = \mu_{\mathrm{if}}(Q') + \sum_{k=1}^{N} \frac{\partial \mu_{\mathrm{if}}}{\partial Q'_k} Q'_k + \frac{1}{2} \sum_{k=1}^{N} \sum_{l=1}^{N} \left( \frac{\partial^2 \mu_{\mathrm{if}}}{\partial Q'_k \partial Q'_l} \right)_0 Q'_k Q'_l + \cdots \quad (3)$$

where $Q'$ is the equilibrium geometry of the final electronic state and $N$ is the number of vibrational normal modes. The zeroth order term in Eq. (3) is a static transition dipole moment, which derives from a direct application of the Franck–Condon approximation. Conversely, the first-order term corresponds to the Herzberg–Teller approximation. By using the bra–ket notations ($|\psi_{\upsilon}\rangle = |\upsilon\rangle$ and $|\psi'_{\upsilon'}\rangle = |\upsilon'\rangle$) to express the vibrational wavefunctions of the initial and final electronic states as eigenstates of the $N$-dimensional harmonic oscillator, Eq. (2) can be rewritten as

$$
\begin{aligned}
\langle \Psi_{\mathrm{i}} \mid \mu \mid \Psi_{\mathrm{f}} \rangle = \mu_{\mathrm{if}}(Q_0') \langle \upsilon | \upsilon' \rangle + \sum_{k=1}^{N} \frac{\partial \mu_{\mathrm{if}}}{\partial Q_k'} \langle \upsilon \mid Q_k' \mid \upsilon' \rangle \\
+ \frac{1}{2} \sum_{k=1}^{N} \sum_{l=1}^{N} \left( \frac{\partial^2 \mu_{\mathrm{if}}}{\partial Q' \partial Q_l'} \right)_0 \langle \upsilon \mid Q_k' Q_l' \mid \upsilon' \rangle
\end{aligned} \tag{4}
$$

Although in principle the vibronic absorption emission spectra can be obtained directly by computing Eq. (4), in practice, additional computational strategies need to be employed to calculate the vibronic spectra of medium-to-large-size molecular systems.[11-14, 16, 39] It should be noted that the vibronic calculations described above provide the nuclear (Franck–Condon) part of Eq. (4), whereas the electronic transition moment in Eq. (3) is taken from the electronic structure calculation of the final state. Since transition moments between two electronic excited states are not available from linear-response TDDFT, the moment used here for an $S_n \rightarrow S_{n'}$ transition is that of the corresponding $S_0 \rightarrow S_{n'}$ transition.[40-41] Consequently, the present scheme does not provide absolute or relative ESA intensities, and we restrict our discussion to the vibronic band shapes and peak positions, which are determined solely by the Franck–Condon factors and the PESs and are independent of the electronic transition moment.

As can be evinced from the data reported in Figure 1, the vibronic transition calculations can be employed to obtain the absorption and emission spectra involving any two electronic states, as long as the states' PESs are well determined.[13-14, 42] The detailed information on the PESs of the said two electronic states, such as the atoms' masses and Cartesian coordinates, the electronic energies, and the vibrational normal modes, is easily obtained by calculating the vibrational frequencies of the equilibrium geometry of a given molecule in its electronic states. After the vibrational frequency calculation of two electronic states has been performed, the calculation of the vibronic transition is conducted to obtain vibronic transition lines. To obtain the absorption and emission spectra, all vibronic transition lines for a given electronic transition are convoluted with a Gaussian envelope with a full-width-half-maximum (FWHM, $\omega_{\mathrm{FWHM}}$),

$$S(\omega) = \sum_i I_i \exp\left( -4\ln(2) \frac{(\omega - E_i + \sigma)^2}{\omega_{\mathrm{FWHM}}{}^2} \right) \tag{5}$$

where $S$ is the calculated spectrum, $I_i$ is intensity of the $i^{\mathrm{th}}$ vibronic line, $E_i$ is the energy of the $i^{\mathrm{th}}$ vibronic transition, and $\sigma$ is a correction factor, which is the value of the frequency difference between the calculated peaks and their experimentally measured counterparts.[43-44] The Gaussian envelopes ($\omega_{\mathrm{FWHM}}$) and correction factors for the absorption and emission transitions were determined by comparing them with the corresponding experimentally measured absorption and emission peaks, as summarized in Table S1. The $\omega_{\mathrm{FWHM}}$ of the ESA transitions ($S_n \rightarrow S_{n'}$) was approximately determined by adding two $\omega_{\mathrm{FWHM}}$'s for the $S_0 \rightarrow S_n$ and $S_0 \rightarrow S_{n'}$ transitions. The value of $\sigma$ of the ESA transition ($S_n \rightarrow S_{n'}$) was approximately determined by calculating the difference of the two $\sigma$ values for the $S_0 \rightarrow S_n$ and $S_0 \rightarrow S_{n'}$ transitions. Note that the $\omega_{\mathrm{FWHM}}$ of the ESA transition in our method is approximated to be estimated by the sum of the line broadening of two relevant transitions. Such an approximation may fail when the charge distributions of electronic states are significantly different. For example, $S_0 \rightarrow S_n$ and $S_0 \rightarrow S_{n'}$ transitions exhibit a large line broadening due to larger charge separation in $S_n$ and $S_{n'}$ compared with $S_0$, but $S_n \rightarrow S_{n'}$ transition could be narrow.

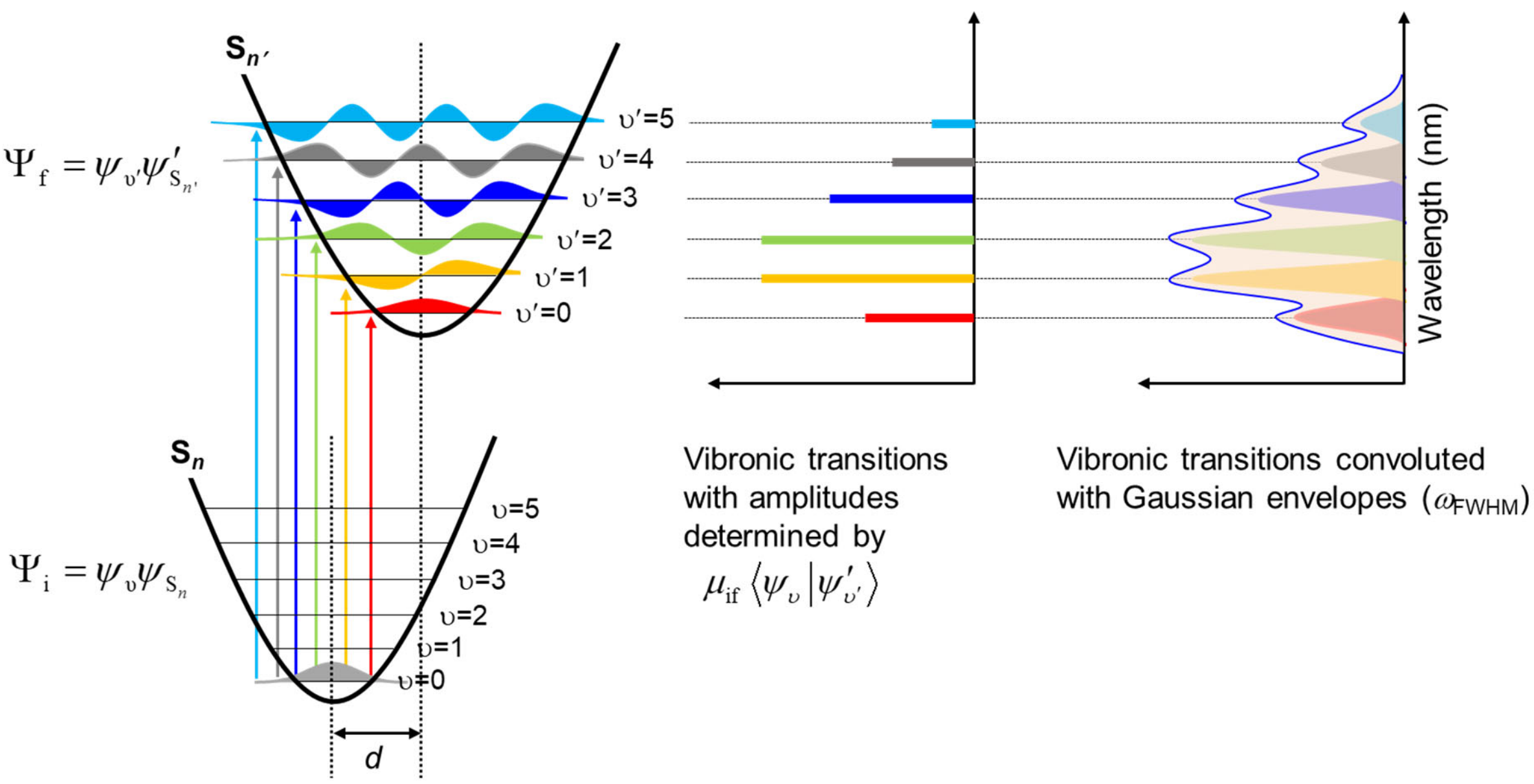

**Figure 1.** Schematic of the calculation of the vibronic transitions ($S_n \rightarrow S_{n'}$).

The molecular geometry optimization, vibrational frequencies, and vibronic spectra were obtained by applying the DFT[45] and TDDFT[46-49] methods at the APFD,[50] B3LYP,[51] CAM-B3LYP,[52] and $\omega$B97XD[53] levels with a 6-311++G(d,p) basis set,[54] as implemented in the Gaussian 16 package.[42] The vibronic transition calculations were carried out by using Franck-Condon-Herzberg-Teller approximation, adiabatic Hessian, and time-independent framework at 298 K. The optimized molecular structures of azulene in different electronic states are reported in Table S2 and Figure S1. The $S_0 \rightarrow S_1$ and $S_0 \rightarrow S_2$ transitions of azulene were calculated using different DFT functionals (APFD, B3LYP, CAM-B3LYP, and $\omega$B97XD) and 6-311++G(d,p) basis set, as can be evinced from Figure S2. Although the 0–0 transition energies in all calculated spectra are not in good agreement with the experimentally determined values, the overall features of the experimental absorption peaks of azulene are well reproduced by the vibronic transition calculations. The calculated vibronic absorption spectra do not depend much on the DFT functionals. Accordingly, all calculation results with APFD/6-311++G(d,p) will be discussed in this work.

## 3. Results and discussion

To demonstrate our practical method for calculating ESA spectra, we used azulene as a model system. Unlike most organic molecules, which obey Kasha's rule,[55] azulene has been shown to exhibit a strong fluorescence from the $S_2$ to the $S_0$ state characterized by a lifetime in the 1–1.6 ns range in different solvents.[56] The lifetime of azulene in the $S_1$ state was determined to be 1.7 ps in cyclohexane of which the fast lifetime has been shown to result from the conical intersection of $S_1$ and $S_0$.[57-58] By selectively exciting azulene to its $S_1$ and $S_2$ states, TA studies have been conducted on azulene by several groups, and the ESA spectra of azulene are found in the literature.[57, 59] The PESs and vibrational normal modes of azulene in different electronic states were obtained by employing DFT and TDDFT methods. Subsequently, the vibronic transition calculations between two electronic states allowed us to obtain the corresponding absorption and emission spectra. To validate our method, the experimental absorption, emission, and ESA spectra of azulene were compared directly with those obtained performing vibronic transition calculations.

### 3.1. Vibronic absorption transitions involving the $S_0$ state of azulene

First, we obtained the absorption spectrum of azulene by calculating the vibronic transitions from the electronic ground state ($S_0$) to higher-energy excited states ($S_n$). The calculated absorption transitions of azulene, reported in Figure 2(a–c), were compared with the experimentally measured ones reported in Figure 2(d–f). Note that the $S_0 \rightarrow S_5$ transition is not reported because the oscillator strength for the $S_0 \rightarrow S_5$ transition was found to be zero. Based on the vibronic transition calculations, the absorption peaks at 700 and 350 nm were readily assigned to the $S_0 \rightarrow S_1$ and $S_0 \rightarrow S_2$ transitions, respectively, and the absorption band at ~280 nm can be attributed to the $S_0 \rightarrow S_3$ and $S_0 \rightarrow S_4$ transitions. The emission spectra were calculated for the transitions from the $S_2$ to the $S_0$ state (Figure 2(g)) and from the $S_1$ to the $S_0$ state (Figure 2(h)). The calculated absorption and emission spectra of azulene are in excellent agreement with their experimentally measured counterparts reported in Figure 2. This observation indicates that the transition energies, PESs, and vibrational normal modes of azulene in different electronic states are reasonably well approximated following implementation of DFT-based methods to calculate the vibronic transitions of absorption and emission spectra.

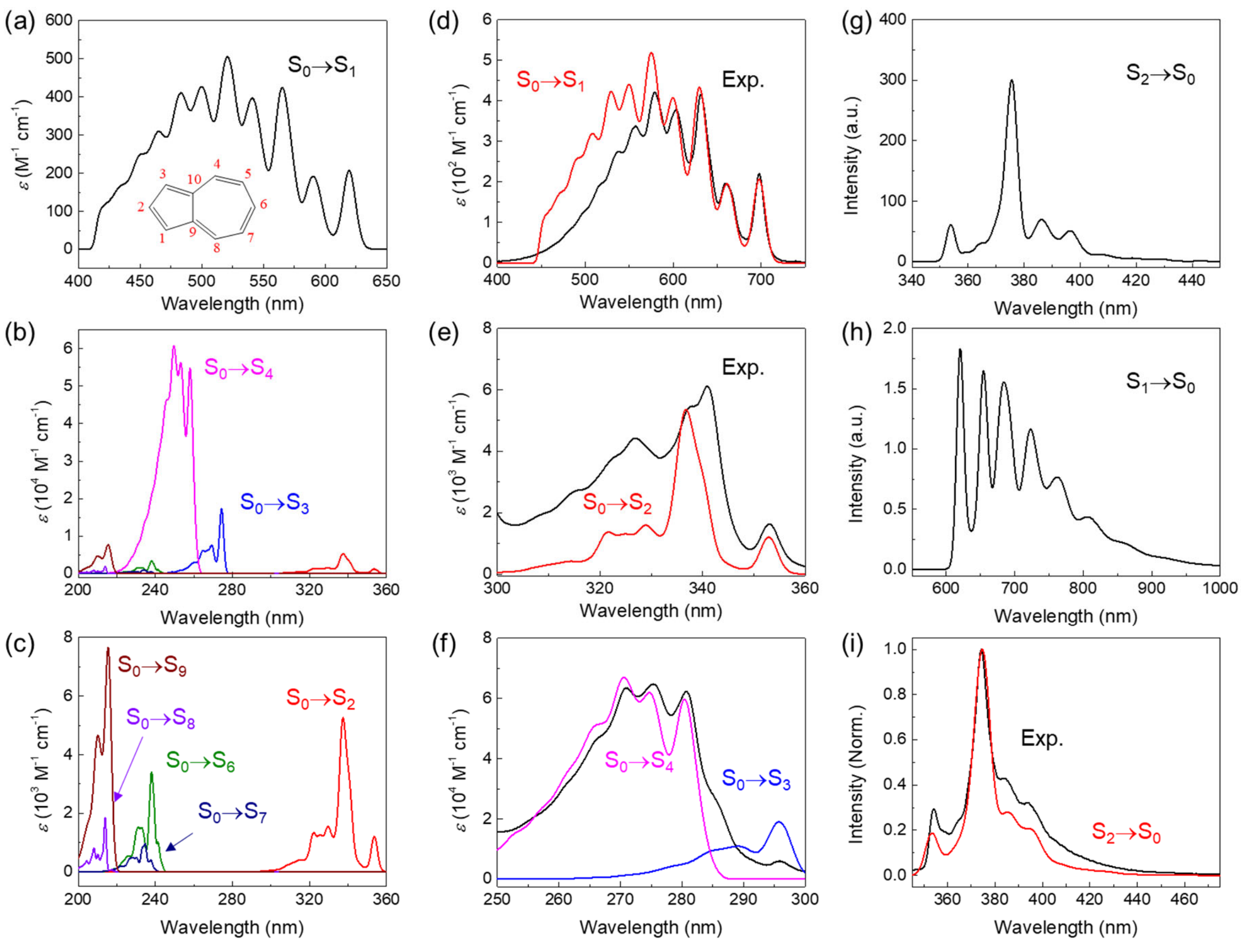

**Figure 2**. Vibronic transition calculations of absorption and emission spectra of azulene. (a), (b), (c), $S_0 \rightarrow S_n$ transitions ($\omega_{FWHM}$ = 300 cm$^{-1}$). Note that the structure of azulene is reported in the inset of (a). (d), (e), and (f) Comparisons between the experimental absorption spectra

of azulene in cyclohexane (black lines) and the calculated absorption transitions. (g) Calculated emission spectrum for the $S_2 \rightarrow S_0$ transition ($\omega_{FWHM}$ = 300 $cm^{-1}$). (h) Calculated emission spectrum for the $S_1 \rightarrow S_0$ transition ($\omega_{FWHM}$ = 300 $cm^{-1}$). (i) Comparison between the experimental fluorescence spectrum of azulene in cyclohexane ($\lambda_{ex}$ = 340 nm) and the calculated emission spectrum.

### 3.2. Vibronic absorption transitions of azulene from $S_1$ and $S_2$ to higher electronic states

TA and PIA signals measured in optical experiments are the result of the sum of GSB, SE, and ESA contributions. The GSB contribution can be obtained by calculating the vibronic transitions from a ground electronic state ($S_0$) to higher-energy electronic states ($S_n$, $n \geq 2$); the SE contribution can be obtained from the calculated vibronic transitions from higher-energy electronic states ($S_n$) to lower-energy electronic states ($S_{n'}$, $n > n'$); the ESA contribution can be obtained from the vibronic transitions from an electronic excited state ($S_{n'}$) to another, higher-energy, electronic excited state ($S_n$, $n > n'$). The GSB and SE contributions are equivalent to the absorption and emission spectra, which are reliably obtained by performing currently available theoretical calculations. However, the ESA contribution is more challenging to calculate. In Figure 3(a, b) are reported our calculated ESA spectra of azulene associated with the transition from the $S_1$ and $S_2$ states to higher-energy excited states, respectively, based on the vibrational transition calculations. Note that the $S_1 \rightarrow S_5$ and $S_2 \rightarrow S_5$ transitions are not reported in Figure 3(a, b), because no intensity could be assigned to them within the present scheme, in which the electronic transition moment is that of the $S_0 \rightarrow S_5$ transition, whose oscillator strength vanishes. This does not imply that the $S_1 \rightarrow S_5$ and $S_2 \rightarrow S_5$ transitions are themselves forbidden. The transition from the $S_2$ state to the $S_1$ state reported in Figure 2(c) corresponds to the SE spectrum.

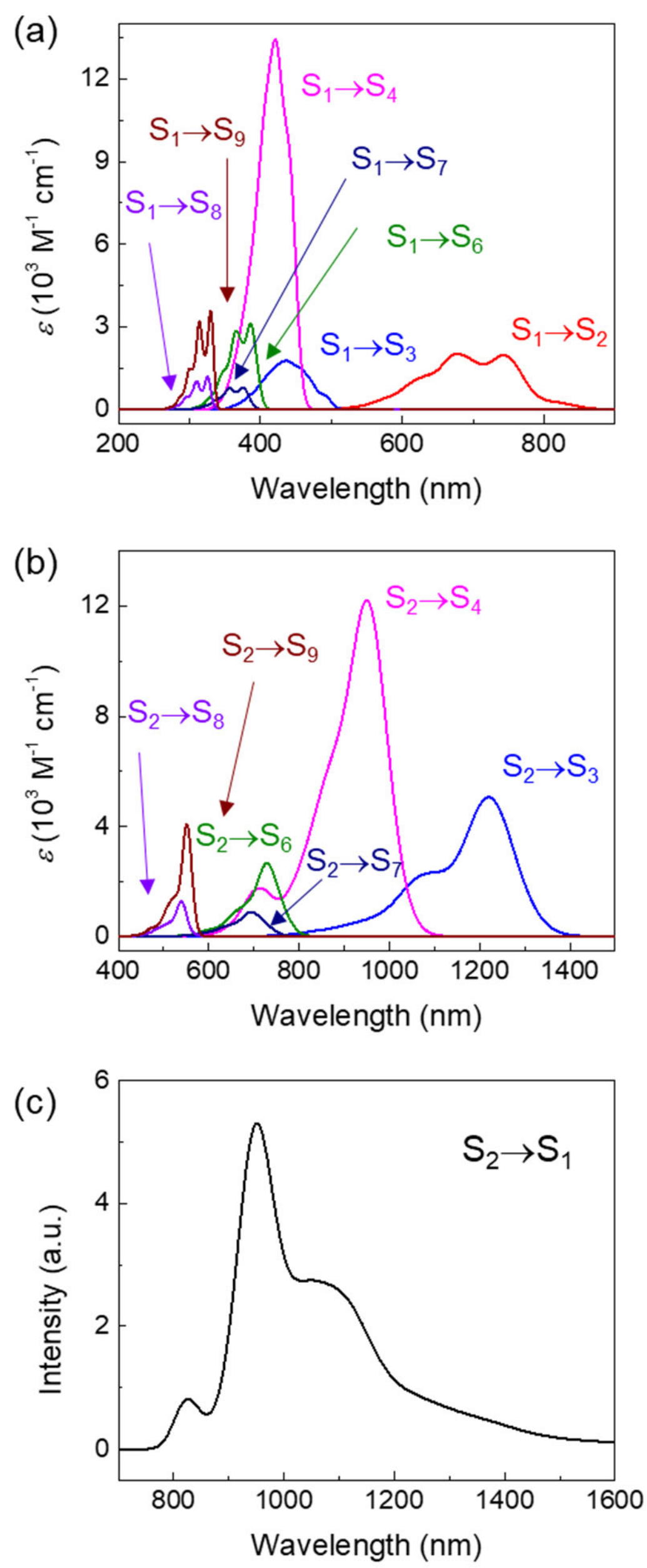

**Figure 3**. Calculated excited-state absorption spectra of azulene. (a) $S_1 \rightarrow S_n$ transitions ($n$ >1). (b) $S_2 \rightarrow S_n$ transitions ($n$ >2). (c) $S_2 \rightarrow S_1$ transition (stimulated emission spectrum). In all cases, $\omega_{FWHM}$ = 800 cm$^{-1}$. The intensities are in arbitrary units and are not comparable between different electronic transitions.

In Figure 4(a) is reported the ESA spectrum of azulene in the $S_1$ state, which was experimentally measured by Foggi and coworkers.[59] These researchers assigned the ESA peaks observed at 450, 400, and 375 nm to the $S_1 \rightarrow S_3$, $S_1 \rightarrow S_6$, and $S_1 \rightarrow S_7$ transitions, respectively, based on their calculations of transition energies and oscillator strengths performed implementing a complete active space multi-configurational self-consistent field (CASSCF) method.[59] However, results from our vibronic transition calculations indicate that the ESA peaks at ~450 and ~375 nm should result from the $S_1 \rightarrow S_3$ and $S_1 \rightarrow S_4$ transitions, respectively, as can be evinced from the data reported in Figure 4(b).

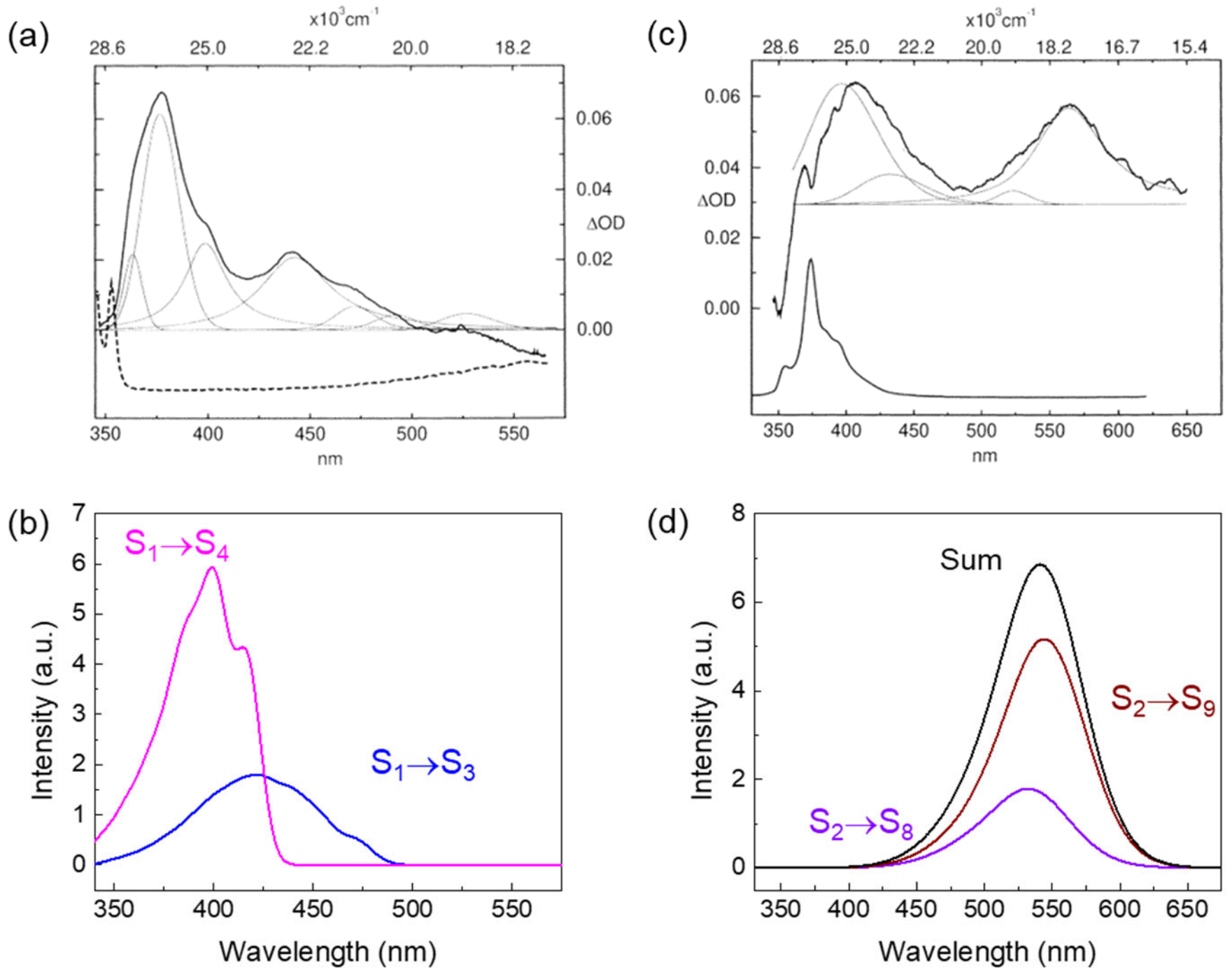

**Figure 4**. (a) Transient absorption (TA) spectrum (thick solid line) of azulene measured following excitation at 680 nm and 0.2 ps delay time. The thin solid lines are a multi-Gaussian fit of the TA spectrum. The dashed line is the ground-state absorption. (b) Our calculated excited-state absorption (ESA) transitions. Note that the calculated intensities are given in arbitrary units. Because the relative intensities of different ESA transitions are not predicted by the present scheme (see Section 2), the calculated $S_1 \rightarrow S_4$ transition is scaled by 1/13 so that its vibronic band shape can be compared with the experimental spectrum. (c) TA spectrum (solid line) of azulene measured by excitation at 340 nm and 15 ps delay time. The thin solid lines are a multi-Gaussian fit of the TA spectrum. The spectrum at the bottom of the graph is a fluorescence spectrum. (d) Our calculated ESA transitions. $\omega_{\mathrm{FWHM}} = 2000\ \mathrm{cm}^{-1}$ in (a) and (c)

In Figure 4(c) is reported the ESA spectrum of azulene in the $S_2$ state, which includes two ESA peaks at ~575 and ~400 nm.[59] Based on the results of CASSCF calculations, the ESA peak at 575 nm had been assigned to the $S_2 \rightarrow S_{11}$ transition.[59] However, results from our vibronic transition calculations indicate that the ESA peak at ~575 nm is due to the $S_2 \rightarrow S_8$ and $S_2 \rightarrow S_9$ transitions. These results demonstrate that vibronic transition calculations reproduce the vibrational band shapes and positions of the ESA features of azulene and thereby provide a useful basis for assigning the observed ESA peaks to specific electronic transitions. We note that the present assignments rest on the transition energies and vibronic band shapes rather than on the calculated intensities..

### 3.3. Vibrational normal modes coupled to electronic transitions

Vibronic transition calculations also provide information on the vibrational normal modes coupled to the corresponding electronic transitions. The calculated absorption spectra associated with azulene's $S_0 \rightarrow S_4$ and $S_1 \rightarrow S_4$ transitions are reported in Figure 5 as an example. Azulene exhibits a planar structure in the $S_0$ and $S_1$ states, whereas its structure is bent in the $S_4$ state. Because the bent structure arises from a distortion along a non-totally symmetric out-of-plane coordinate, the corresponding PES is expected to be a double well and thus anharmonic; the harmonic adiabatic-Hessian treatment used here therefore describes this coordinate only approximately, and the following assignment should be regarded as qualitative. As a consequence, the molecular structure of azulene switches from planar to bent when azulene undergoes excitation from $S_0$ to $S_4$ or from $S_1$ to $S_4$. For such a structural change to take place, the out-of-plane bending vibrational modes of azulene should be mainly driven by and coupled to the electronic transition. Performing the vibrational transition calculations enables us to identify the out-of-plane bending mode ($\nu_2$, 102 cm$^{-1}$), which is mainly coupled to the $S_0 \rightarrow S_4$ and $S_1 \rightarrow S_4$ transitions, as can be evinced from Figure 5. In addition, the evidence reported clearly indicates that the combination of $\nu_2$ and $\nu_{12}$ (675 cm$^{-1}$) is coupled to the $S_0 \rightarrow S_4$ transition and that the combination of $\nu_2$ and $\nu_{18}$ (851 cm$^{-1}$) is coupled to the $S_1 \rightarrow S_4$ transition. Therefore, the vibronic transition calculations are also useful for the direct identification of the vibrational normal modes coupled to a given electronic transition, which are driven for a molecule to undergo a structural change from an initial electronic state to a final electronic state.

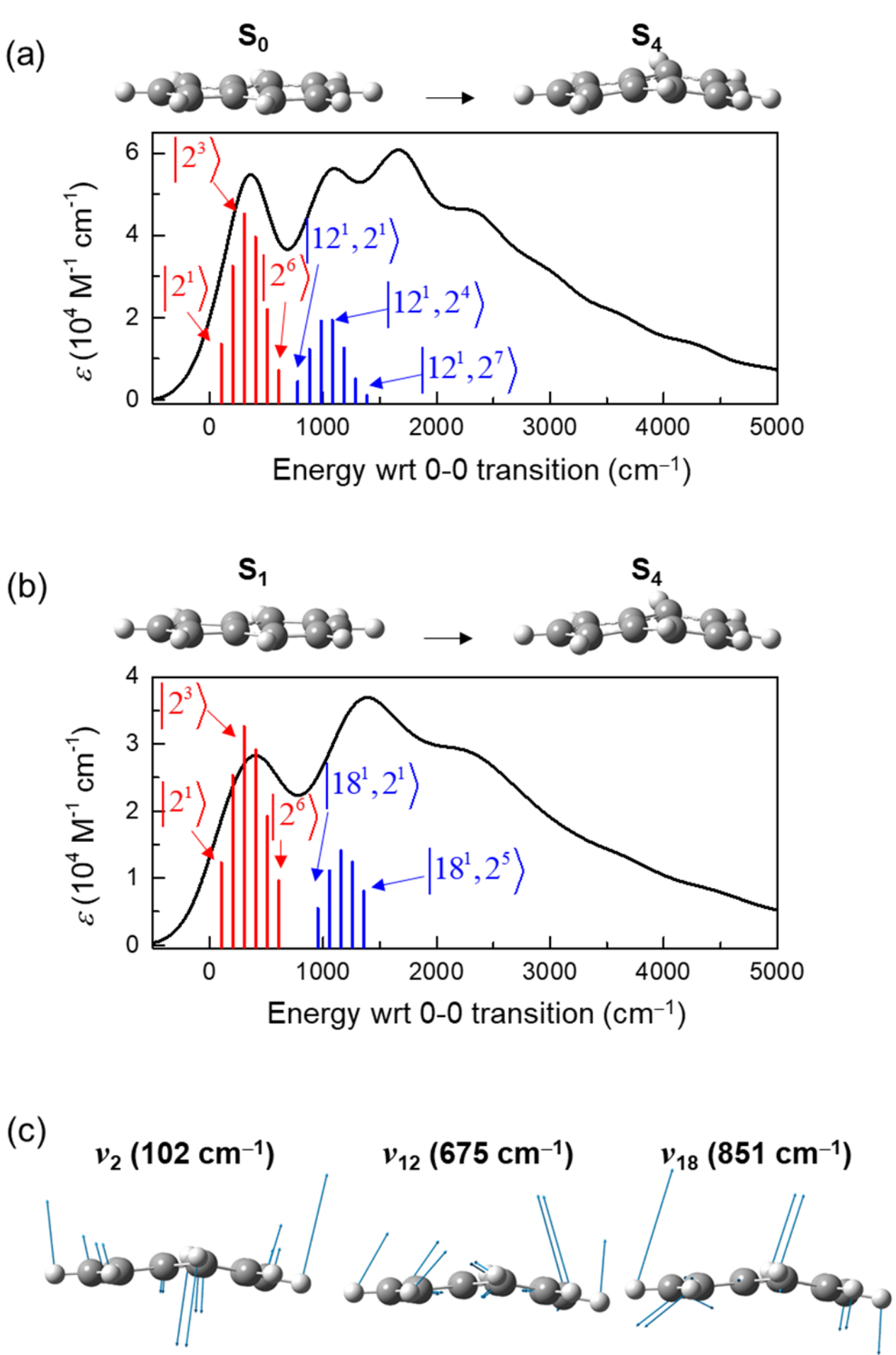

**Figure 5**. (a) Absorption spectrum associated with an azulene $S_0 \rightarrow S_4$ transition and (b) an azulene $S_1 \rightarrow S_4$ transition. The vibronic transitions are assigned to $|N^{\upsilon}\rangle$ from the vibrational ground state of the initial electronic state, where $N$ is the number of vibrational normal modes of the final electronic state and $\upsilon$ is its quantum number. (c) The vibrational normal modes of the final electronic state. Displacement vectors are indicated by the blue arrows.

**4. Concluding remarks**

In this work, we have demonstrated that the vibronic transition calculations can be effectively used to obtain azulene's TA and PIA spectra consisting of the GSB, SE, and ESA contributions as well as absorption and emission spectra based on the PESs of electronic states and all vibrational frequencies, which can be readily obtained by employing DFT and TDDFT methods. Vibronic transition calculations allow straightforward assignment of the electronic transitions as a result of the transition's characteristic vibronic features. However, the values for the transition energies calculated by our method are not as accurate as those calculated *via* other methods and are dependent on the DFT functionals. Conversely, the highest computational cost in our method is associated with the vibrational frequency calculation in each electronic state, particularly in excited states. Nowadays, analytic vibrational frequency calculations can be reliably carried out at the TDDFT level.[47-49]

Like other methods, our method has limitations resulting from the TDDFT method and the harmonic approximation. The PESs of the electronic states with very high energies cannot be accurately determined by the TDDFT method. The harmonic approximation cannot be used when the PESs are highly anharmonic.[60] In addition, the present scheme does not provide transition moments between two excited states, and the calculated ESA intensities therefore carry the electronic transition moment of the corresponding $S_0 \rightarrow S_{n'}$ transition. Absolute and relative ESA intensities would require quadratic-response TDDFT or an explicit evaluation of excited-to-excited transition moments from transition densities.[40] Accordingly, this practical method, requiring only standard DFT and TDDFT calculations, provides an accessible route to interpret the ESA bands of the TA and PIA spectra between low-lying excited states of molecules..

**Acknowledgments**

This study was supported by grants from the National Research Foundation of Korea (NRF) funded by the Korean government (MSIP) (RS-2022-NR069080 for S. P. and RS-2026-25484674 for J. F. J.)

**Declaration of competing interest**

Authors declares no conflict of interest.